\documentclass[aps,pre,showpacs,twocolumn,superscriptaddress]{revtex4}

\usepackage{epsfig}
\usepackage{amsmath,amssymb,wasysym,epsfig,capt-of,ifthen,calc}
\usepackage{bbm}
\usepackage{latexsym}   
\usepackage{setspace}   
\usepackage{array}      
\usepackage{delarray}   
\usepackage{afterpage}
\usepackage{graphicx}
\usepackage{dcolumn}
\usepackage{bm}
\usepackage{array}
\usepackage{hyperref}
\usepackage{float}
\usepackage{supertabular}
\usepackage{longtable}
\newcommand{\be}{\begin{equation}}
\newcommand{\ee}{\end{equation}}
\newcommand{\bea}{\begin{eqnarray}}
\newcommand{\eea}{\end{eqnarray}}

\bibstyle{apsrev}

\begin{document}

\title{Random Sampling vs. Exact Enumeration of Attractors in Random Boolean Networks}
\author{Andrew Berdahl}
 \affiliation{Complexity Science Group, Department of Physics \& Astronomy, University of Calgary, Canada}
\author{Amer Shreim}
 \affiliation{Complexity Science Group, Department of Physics \& Astronomy, University of Calgary, Canada}
\author{Vishal Sood}
 \affiliation{Complexity Science Group, Department of Physics \& Astronomy, University of Calgary, Canada}
\author{Maya Paczuski}
 \affiliation{Complexity Science Group, Department of Physics \& Astronomy, University of Calgary, Canada}
\author{J\"orn Davidsen }
 \affiliation{Complexity Science Group, Department of Physics \& Astronomy, University of Calgary, Canada}

\date{\today}

\begin{abstract}

  We clarify the effect different sampling methods and weighting
  schemes have on the statistics of attractors in ensembles of random
  Boolean networks (RBNs).  We directly measure cycle lengths of
  attractors and sizes of basins of attraction in RBNs using exact
  enumeration of the state space.  In general, the distribution of
  attractor lengths differs markedly from that obtained by randomly
  choosing an initial state and following the dynamics to reach an
  attractor.  Our results indicate that the former distribution decays
  as a power-law with exponent $1$ for all connectivities $K>1$ in the
  infinite system size limit. In contrast, the latter distribution
  decays as a power law only for $K=2$.  This is because the mean
  basin size grows linearly with the attractor cycle length for $K>2$,
  and is statistically independent of the cycle length for $K=2$. We
  also find that the histograms of basin sizes are strongly peaked at
  integer multiples of powers of two for $K<3$.

\end{abstract}


\maketitle
\section{Introduction}

Random Boolean Networks (RBNs)~\cite{Kauffman1969} have been widely
used as elementary models for genetic regulation. In such a model, a
binary state based on Boolean
logic~\cite{DrosselRBNreview2008,bornholdt05,aldanaRBN2003}
encapsulates local gene expression. An RBN consists of $N$ Boolean
(0,1) elements where the value of each element evolves in discrete
time according to a random Boolean function of $K$ randomly chosen
distinct inputs.  Within an annealed
approximation~\cite{DerridaPomeau}, the RBN can be in one of three
phases: a frozen phase ($K=1$), in which a perturbation to the state
of a single node can propagate only to a finite number of nodes; a
chaotic phase ($K>2$), in which the perturbation spreads to a finite
fraction of the nodes; and a critical phase ($K=2$), which lies in
between the frozen and chaotic phases.

In a finite RBN the number of possible states is also finite.
Therefore, each dynamical state in a deterministically updated RBN is
either transient or belongs to an attractor cycle. A transient state
may be reached no more than once during the dynamics. Meanwhile, a
state which belongs to an attractor cycle may be reached infinitely
often if the chosen initial state falls into the basin of attraction
of the attractor cycle. Many analytical results concerning the
duration of transients and attractor cycle lengths have been
established for the random map (RM), which is the limit of RBNs when $K \to
N$.  These analytical arguments are based on the fact that the
annealed approximation is exact for the random
map~\cite{DerridaFlyvbjerg,bastollaparisi}. However, most of the known
results for $K=2$ critical RBNs come from numerical simulations.

Since the size of the state space grows as $2^N$, an exhaustive search
of attractors is infeasible for large $N$.  Instead, a random sampling
procedure has (almost exclusively) been
used~\cite{Kauffman1969,bhattaPRL,paczuskiRBN}: the system is prepared
in a randomly chosen initial state, and the RBN rules are used to
evolve the system until an attractor is reached. Typically a fixed
number of initial conditions are used for each realization of an RBN,
and the ensemble properties are obtained by repeating this procedure
for many realizations.  Using this sampling procedure, it has been
shown that the lengths of the transients to reach an attractor, as
well as the lengths of the attractor cycles reached are both power-law
distributed for the ensemble of $K=2$ critical RBNs~\cite{bhattaPRL}.
However, the size of the state space limits the accuracy of the random
sampling procedure. For instance, small basins of attraction may be
under-sampled as suggested by recent studies of the (mean) number of
attractors~\cite{samuelsson2003RBN,BilkeEnum2001,Drossel2005PRL}.
This has raised strong concerns regarding the validity of estimates of
the distributions of cycle lengths and sizes of basins of attraction
that are based on the random sampling procedure~\cite{aldanaRBN2003}
--- like the ones obtained in Ref.~\cite{bhattaPRL}.

In particular, the simple sampling procedure described above can only measure
the distribution of attractor lengths \emph{reached from a randomly chosen
  initial state}, but not the unbiased distribution of attractor cycle lengths
itself. These two distributions only coincide if there are no correlations
between the length of an attractor cycle and its basin size.  While it can be
shown analytically that the unbiased distribution of attractor cycle lengths
decays as a power-law with exponent $1$ for the random
map~\cite{aldanaRBN2003}, less is known about $K<N$.
Refs.~\cite{ShmulevichEntropyRBNPRE,ShmulevichEntropyRBNPRL} estimate a basin
entropy using exact enumeration for various $K$, but do not consider attractor
cycle lengths.

In this article, we clarify the effects of different sampling
procedures and weighting schemes by making exact enumerations of the
state space of RBNs to estimate the distributions of both the sizes of
the basins of attraction and the attractor cycle lengths.  By
weighting each attractor by its basin size, we can reproduce the
distributions obtained by randomly sampling initial states as
discussed above.  We find numerically that the unbiased distribution of
attractor cycle lengths differs markedly from that obtained by
randomly sampling initial states for all $K>2$. This is corroborated
by analytical arguments based on an annealed
approximation~\cite{bastollaparisi}. Remarkably, for $K=2$ both
distributions are well approximated by the same power-law. This
difference between critical and chaotic RBNs is related to how the
basin size depends on the length of its attractor cycle. We show
analytically that the mean basin size increases linearly with the
length of the attractor for the random map. Numerically, we find that
this also holds for $K>2$. For $K=2$, the mean basin size is
independent of the attractor cycle length.  Enumeration also allows us
to study the distribution of basin sizes; we find that for $K=2$,
those distributions are discrete and strongly peaked at integer multiples
of powers of two, reflecting in part the analytical structure
previously found for $K=1$~\cite{flyvbjerg1988esk}.


In Section II, we present a physical motivation for the different
weighting schemes used to compute various distributions for cycle
lengths or basin sizes as well as a more formal discussion clarifying
the mathematical relations between these distributions. Section III
contains the results from numerical simulations, while Section IV
concludes with a summary of the main findings.

\section{Definition of distributions}

Depending on the weighting scheme, different distributions for the
same quantity can be obtained. One can, for instance, make an
estimate of the distribution of attractor lengths that would be
obtained by randomly sampling $I$ initial states (for each of $R$
realizations of the RBN) and following the dynamics to reach an
attractor as follows: For an RBN, make an exhaustive list of each
attractor with its cycle length and size of its basin. Then pick $I$
attractors randomly -- each with a weight proportional to its basin
size -- and compute a histogram of attractor cycle lengths for the
RBN. Repeat this for $R$ different realizations of the RBN and average
the results to obtain an ensemble averaged probability distribution.

The distribution of cycle lengths obtained this way corresponds to an
estimate of $Q_u(l)$ in the notation below.  Note that using a weight
proportional to the basin size accounts for the fact that initial
states have a proportionally higher probability to fall in larger
basins within the state space than in smaller ones.  For simplicity,
we will refer to $Q_u(l)$ as the distribution of cycle lengths
obtained by random sampling -- even though it oversimplifies the
situation. In particular, as this discussion implies, we are not using
a random sampling method, but rather reproducing what would be
obtained using that method
\footnote{In principle, one could use a
  random sampling method to reproduce all the other distributions
  mentioned below by appropriate (re)weighting, e.g. with inverse
  basin size, etc., if the relevant quantities were measured.}.

On the other hand, using the exhaustive list of each attractor and its
basin size for all $R$ realizations of an RBN, one could compute a
histogram by directly accumulating the results for each attractor on
the list -- independent of its basin size and also independent of
which RBN realization it appears in within the ensemble of $R$
realizations.  Such a distribution of attractor lengths corresponds to
$P_{a}(l)$ in the notation below.  In order to assist the general
reader, we will refer to $P_{a}(l)$ as the distribution of cycle lengths
obtained by exact enumeration -- again, even though it oversimplifies
the true situation.

These two distributions differ in two ways as signified by the $P$
vs. $Q$ label as well as the subscript $a$ vs. $u$.  In the notation
we use, $Q$ denotes distributions obtained by weighting attractors
according to the size of their basin of attraction, while $P$ denotes
distributions obtained without regard to the basin size.  In addition,
the subscript $u$ in $Q_u(l)$ denotes a distribution obtained by
weighting each RBN equally (e.g. making a histogram for each RBN and
then uniformly averaging the results over different realizations),
while the subscript $a$ in $P_a(l)$ denotes a distribution in this
case obtained by weighting each attractor equally -- regardless of
which RBN it came from (see below for further discussion).  It is
obvious that there are many other distributions that can be estimated,
depending on the weighting scheme.  In general, there is no reason
these distributions should be similar. However, there are
mathematical relations connecting them as described next.

\subsection{Formal development}

The state space of a single realization of an RBN may contain many
different attractors. Each attractor is characterized by $(l,b)$,
which are its cycle length $l$ and the size $b$ of its basin of
attraction.  For RBN $i$ in a given ensemble, we count each attractor
$\alpha$ in its state space, and record the respective $l$ and
$b$. This allows us to obtain the probability that a randomly chosen
\emph{attractor} of RBN $i$ has cycle length $l$ and basin size $b$,
\begin{eqnarray}
    \label{singleRBNdist}
        P^{(i)}(l,b)
        &\equiv& \frac{1}{A_{i}}\sum_{\alpha=1}^{A_i} \delta_{l_{\alpha},l} \delta_{b_{\alpha}, b} \nonumber\\
        &=&\frac{A_{i}(l, b)}{A_{i}}.
\end{eqnarray}
Here $A_{i}(l,b)$ is the number of attractors with cycle length $l$ and basin
size $b$ in RBN $i$, and $A_i$ is the total number of attractors in it. The
probability that a randomly chosen state lies in the basin of an
attractor with $(l,b)$ for RBN $i$ is given by
\begin{eqnarray}
    Q^{(i)}(l,b)
    &\equiv& \sum_{\alpha=1}^{A_i}\frac{b_\alpha}{2^N}\delta_{l_\alpha,l}\delta_{b_\alpha,b}\nonumber\\
      &=& \frac{b}{2^N} A_{i}(l, b) = \frac{b}{2^N} A_{i} P^{(i)}(l,b). \label{samOneRBN}
\end{eqnarray}
It is easy to check that both $P^{(i)}(l,b)$ and $ Q^{(i)}(l,b)$ are normalized to unity.

To obtain the corresponding probabilities for attractor lengths
and basin sizes for an \emph{ensemble} of RBNs, we assign a
normalized weight $w_i$ to each RBN $i$.  If $R$ is the number
of RBNs in the ensemble, the probability that a randomly chosen
attractor from the ensemble has cycle length $l$ and basin
size $b$, is
\begin{eqnarray}
    P(l, b;\{w_i\}) &\equiv& \sum_{i=1}^{R} w_i P^{(i)}(l,b).
    \label{eelb0}
\end{eqnarray}
Similarly, the probability over an ensemble of RBNs that a randomly chosen
state lies in the basin of an attractor with $(l,b)$ is given by
\begin{eqnarray}
    Q(l,b; \{w_i\}) &\equiv& \sum_{i=1}^R w_i Q^{(i)}(l,b)\nonumber\\
        &=& \frac{b}{2^N}P(l,b; \{w_i A_i\}).
    \label{lbsam}
\end{eqnarray}

For uniform weights $w_i=1/R$, we obtain the distribution,
\begin{eqnarray}
    P_u(l,b) \equiv \frac{1}{R}\sum_{i=1}^R \frac{A_{i}(l, b)}{A_i}. \label{lbdistuni}
\end{eqnarray}

If, in contrast, we use the weights
$w_i = A_i/(R \langle A \rangle$), we obtain a distribution for the ensemble where all attractor
are counted equally:
\begin{eqnarray}
    P_{a}(l,b) \equiv \frac{1}{R \langle A \rangle}\sum_{i=1}^R A_{i}(l,b), \label{lbdistatt}
\end{eqnarray}
where $\langle A \rangle$ is the mean number of attractors in a single
realization of an RBN. 

Obtaining ensemble averaged $"Q"$ distributions proceeds in the same manner.
For uniform weights $w_i=1/R$, Eq.~(\ref{lbsam}) becomes
\begin{eqnarray}
    Q_{u}(l,b)
    &\equiv&\frac{b}{2^N}\frac{1}{R}\sum_{i=1}^R  A_i(l,b)\nonumber\\
    &=& \frac{b}{2^N}\langle A \rangle P_{a}(l,b)
    \label{lbsamuni}
\end{eqnarray}
where we used Eq.~(\ref{lbdistatt}) to obtain the second equality.

For comparisons, we also consider Eq.~(\ref{lbsam}) when the RBNs are weighted
by the inverse of the number of attractors,
\begin{eqnarray}
    Q_{1/a}(l,b)&\equiv& Q(l,b;\{A_i^{-1}/ (R \langle A^{-1} \rangle) \})\nonumber\\
    &=&\frac{1}{R \langle A^{-1} \rangle}
        \frac{b}{2^N}\sum_{i=1}^R \frac{A_i(l,b)}{A_i}\nonumber\\
    &=& \frac{1}{\langle A^{-1} \rangle}  \frac{b}{2^N}P_{u}(l,b).
    \label{lbsamatt}
\end{eqnarray}
Eqs.~(\ref{lbdistuni},\ref{lbdistatt},\ref{lbsamuni},\&
\ref{lbsamatt}) make up the four weighting schemes we focus on in the remainder
of this paper.

Using these joint distributions, we can construct the
distributions of attractor lengths, by summing over the basin sizes $b$.  For
example,
\begin{eqnarray}
  \label{eq:samALenu}
  Q_{u}(l) &\equiv& \sum_b Q_{u}(l,b) = \frac{\langle A \rangle}{2^N}\sum_b b P_{a}(l,b)\nonumber\\
  &=& \frac{\langle A \rangle}{2^N} \langle b(l) \rangle_a P_{a}(l).
\end{eqnarray}
Here we have used Eq.~(\ref{lbsamuni}) and defined the mean basin size of
attractors of length $l$,
\begin{eqnarray}
  \label{eq:meanbasin}
  \langle b(l) \rangle_a \equiv \frac{\sum_b b P_{a}(l,b)}{P_{a}(l)},
\end{eqnarray}
with
\begin{eqnarray}
  \label{eq:samALenu2}
  P_{a}(l) &\equiv& \sum_b P_a (l,b) \nonumber\\
\end{eqnarray}

The distributions $P_{u}(l)$ and $Q_{1/a}(l)$ are defined
analogously. In order to assist the general reader, we refer to $Q_u(l)$ as the
distribution of cycle lengths obtained by random sampling and $P_a(l)$
as the distribution obtained by exact enumeration.

In this paper, all these quantities are estimated by exact enumeration of the
state space. For each realization of the RBN, we find the image of each of the
$2^N$ states under the dynamical map. We connect each state and its image by a
directed link to form the state space network (SSN).  We follow these directed
links to reach the attractors, which are cycles of directed links. For each
attractor we find its cycle length and basin size~\cite{ShreimRBN2008}.

\begin{figure*}
\centering
\begin{tabular}{cc}
\epsfig{file=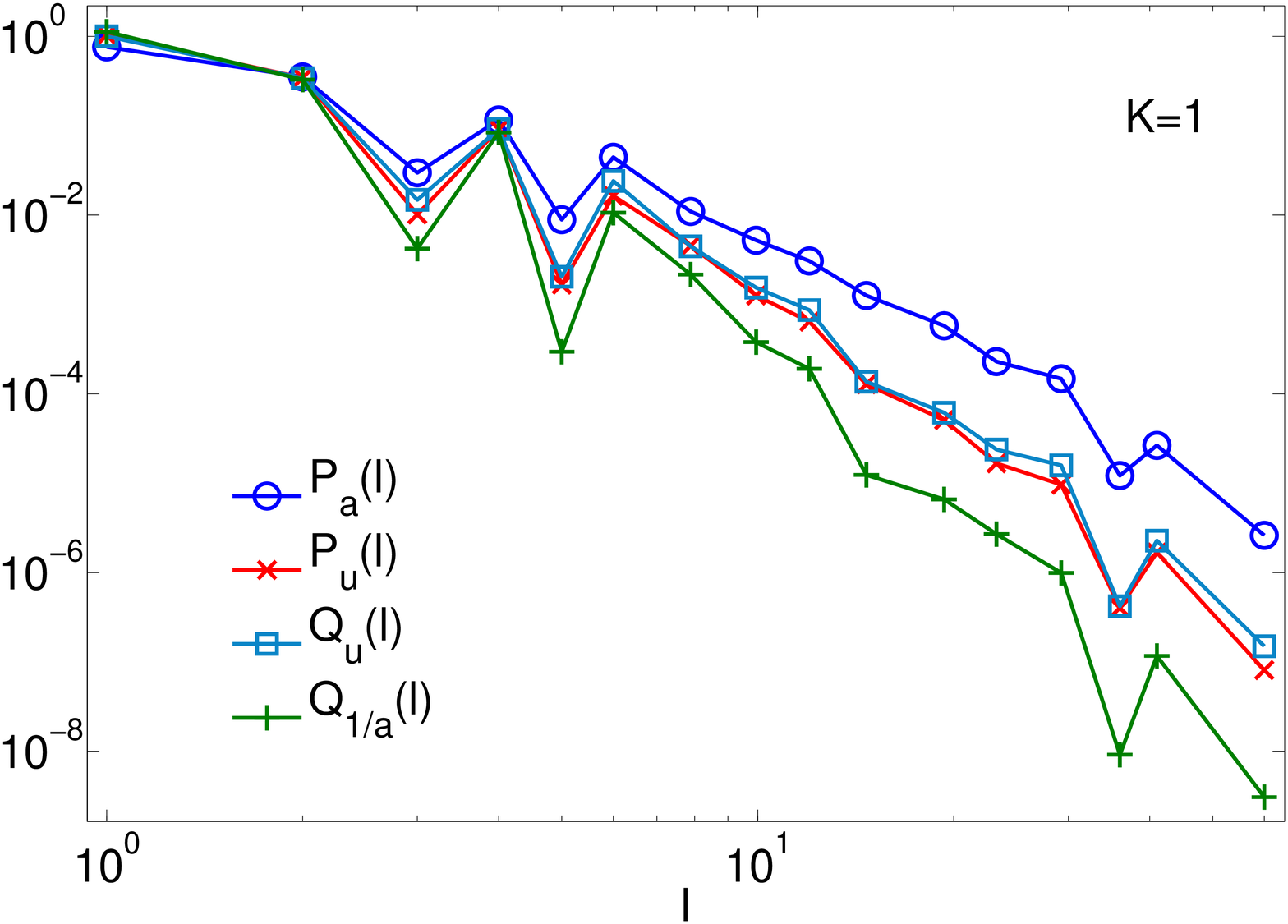,width=0.5\linewidth,clip=} \\
\epsfig{file=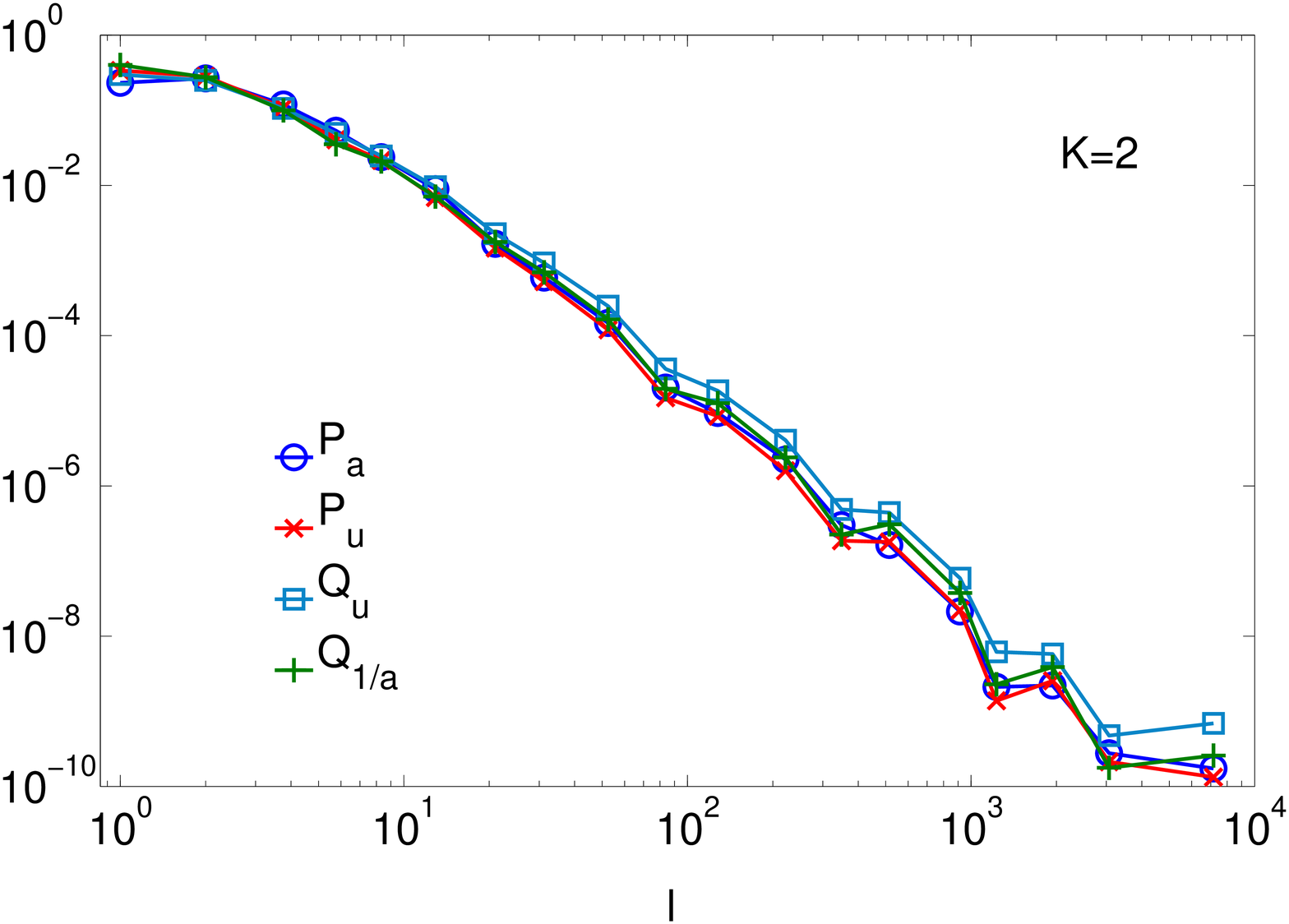,width=0.5\linewidth,clip=} \\
\epsfig{file=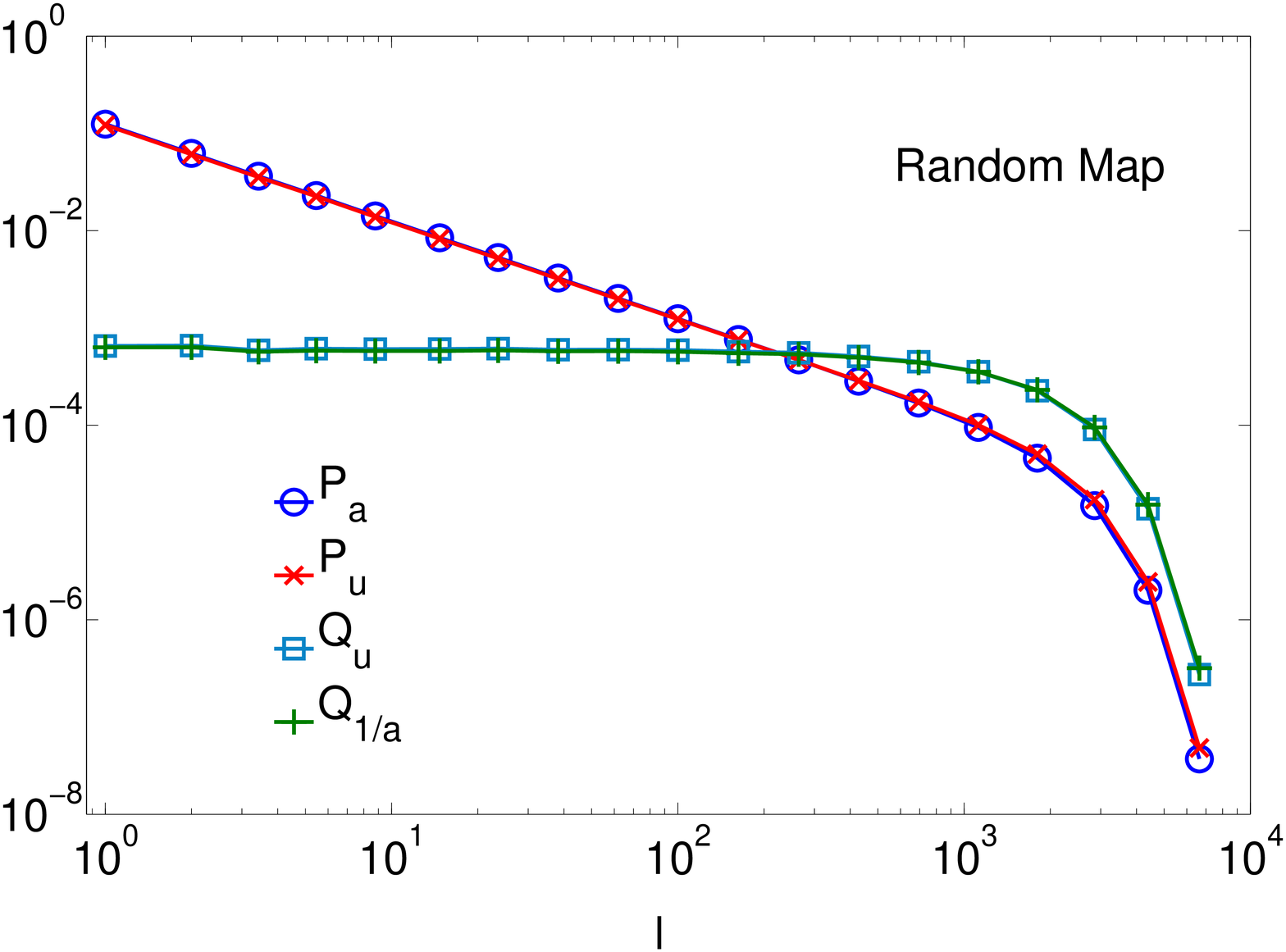,width=0.5\linewidth,clip=}
\end{tabular}
\caption{(Color online.) Distribution of attractor cycle lengths using four
  weighting schemes for $K=1$, $K=2$ and the random map. Results are
  statistically indistinguishable for $K=2$, while differences appear for all
  other $K$.  $Q_u(l)$ represents the (simplest) distribution of cycle lengths
  obtained by random sampling, while $P_a(l)$ represents the (simplest)
  distribution obtained by exact enumeration -- as explained in Section II.
  This and all subsequent figures are based on $5\times10^5$ RBN realizations
  with $N=22$, unless stated otherwise.}
\label{Wschemes}
\end{figure*}


\section{Results}

Fig.~\ref{Wschemes} shows the distribution of attractor lengths
obtained by random sampling $Q_{u}(l)$, the distribution obtained by
exact enumeration $P_{a}(l)$, as well as $P_u(l)$ and $Q_{1/a}(l)$ for
$K=1$, $K=2$ and the random map. Results for other values of $K$ were
also obtained and some of these results are shown later.  We find that
for all $K\neq 1$, $P_{u}(l)$ is similar to $P_{a}(l)$, while
$Q_{u}(l)$ is similar to $Q_{1/a}(l)$. Defining $A_i(l) \equiv \sum_b
A_i(l,b)$, these results imply that the ratio $A_i(l)/A_i$ is
statistically independent of the total number of attractors in the
RBN, $A_i$, for $K\neq 1$. This follows from comparing
Eqs.~(\ref{lbdistuni}) and~(\ref{lbdistatt}) or their analogs for
$Q_u(l)$ and $Q_{1/a}(l)$, respectively. Such independence is expected
for the random map and for $K>2$ in the thermodynamic limit, based on
the annealed approximation~\cite{bastollaparisi}.  We currently have
no explanation for why this would also hold for $K=2$.  For $K=1$, the
difference between $P_u(l)$ and $P_a(l)$ or $Q_{u}(l)$ and
$Q_{1/a}(l)$ is a direct indication that there are significant
correlations between $A_i(l)/A_i$ and $A_i$. This is expected since
long attractors arise from long loops in the network of RBN elements,
which also give rise to a larger total number of
attractors~\cite{flyvbjerg1988esk,Drossel2005PRL}. Since we are mostly
concerned with $K>1$ in what follows, we will focus on the
distribution of cycle lengths obtained by exact enumeration,
$P_{a}(l)$, and the distribution of cycle lengths obtained by randomly
sampling, $Q_{u}(l)$, which differ from each other for all $K\neq 2$
as shown in Fig.~\ref{Wschemes}.

While this difference is well-known for the random map, the relationship between
these two distributions has not been clarified for other values of $K$. As
shown for the random map in Refs.~\cite{harris1960,DerridaFlyvbjerg},
\begin{eqnarray}
  Q_u (l) &=& \sum_{t \geq l} \frac{(2^N - 1)
  !}{(2^N - t) ! (2^N)^t} \nonumber \\
  &\approx& \frac{1}{2^{N / 2}} \int_{l / 2^{N / 2}}^{2^{N / 2}} d y e^{- y^2 /
  2} , \label{RSALSA}
\end{eqnarray}
while in Refs.~\cite{flyvbjerg1988esk,aldanaRBN2003} it was found that
\begin{eqnarray}
  P_a (l) &\approx& \frac{e^{- l^2 / 2^N}}{\langle A \rangle l}.
 \label{EEAL1}
\end{eqnarray}
The latter distribution decays as a power-law with exponent $1$, up to an $N$
dependent cut-off, while the former -- being the complement of the error
function -- is a flat distribution up to an $N$ dependent cut-off.  Indeed,
using the finite size scaling method for different $N$ allows one to collapse
the different curves for $P_{a}(l)$ and $Q_{u}(l)$, respectively.  As
discussed in Section II, $Q_{u}(l)$ is related to $P_{a}(l)$ by a factor which
is the mean basin size of attractors of length $l$, $\langle b(l) \rangle_a$,
see Eq.~(\ref{eq:samALenu}). Hence $Q_{u}(l)$ and $P_{a}(l)$ can only have
different functional forms if $\langle b(l) \rangle_a$ varies with $l$.

\begin{figure}
\centering
\begin{tabular}{cc}
\epsfig{file=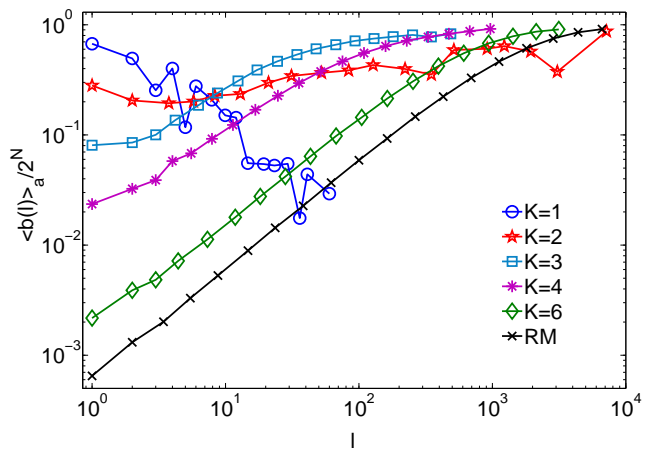,width=1.0\linewidth,clip=}
\end{tabular}
\caption{(Color online.) Mean (normalized) basin size, $\langle b(l)
  \rangle_a/2^N$, vs. attractor length, $l$, for various values of
  $K$ and the random map (RM). $\langle b(l) \rangle_a$ decreases with $l$ for
  $K=1$, and increases for $K>2$ and the random map. For $K=2$, the mean basin
  size is approximately independent of $l$.  Note that the estimated mean
  basin size is only shown for those bins which had at
  least 50 data points.}
\label{meanB}
\end{figure}

\begin{figure}
 \centering
 \epsfig{file=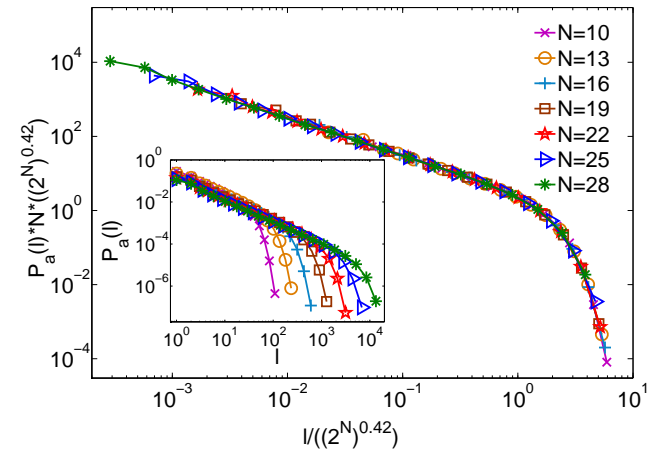,width=1\linewidth,clip=} \\
\begin{center}
(a)
\end{center}
\epsfig{file=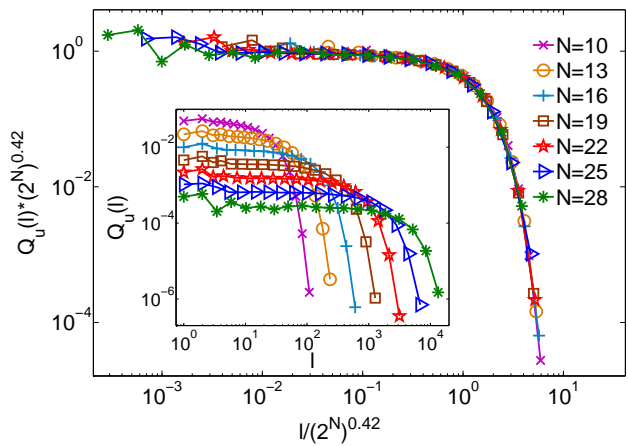,width=1\linewidth,clip=}
\begin{center}
(b)
\end{center}
\caption{(Color online.) Finite size scaling analysis of attractor
  length distributions for $K=6$ RBNs, (a) $P_{a}(l)$ and (b)
  $Q_{u}(l)$.  The insets show the original distributions.  These
  results are based on the following number of realizations:
  $5\times10^5$ for $N=10-22$, $5\times10^4$ for $N=25$ and $5\times
  10^3$ for $N=28$.}
 \label{LaK6}
 \end{figure}

\begin{figure}
 \centering
 \epsfig{file=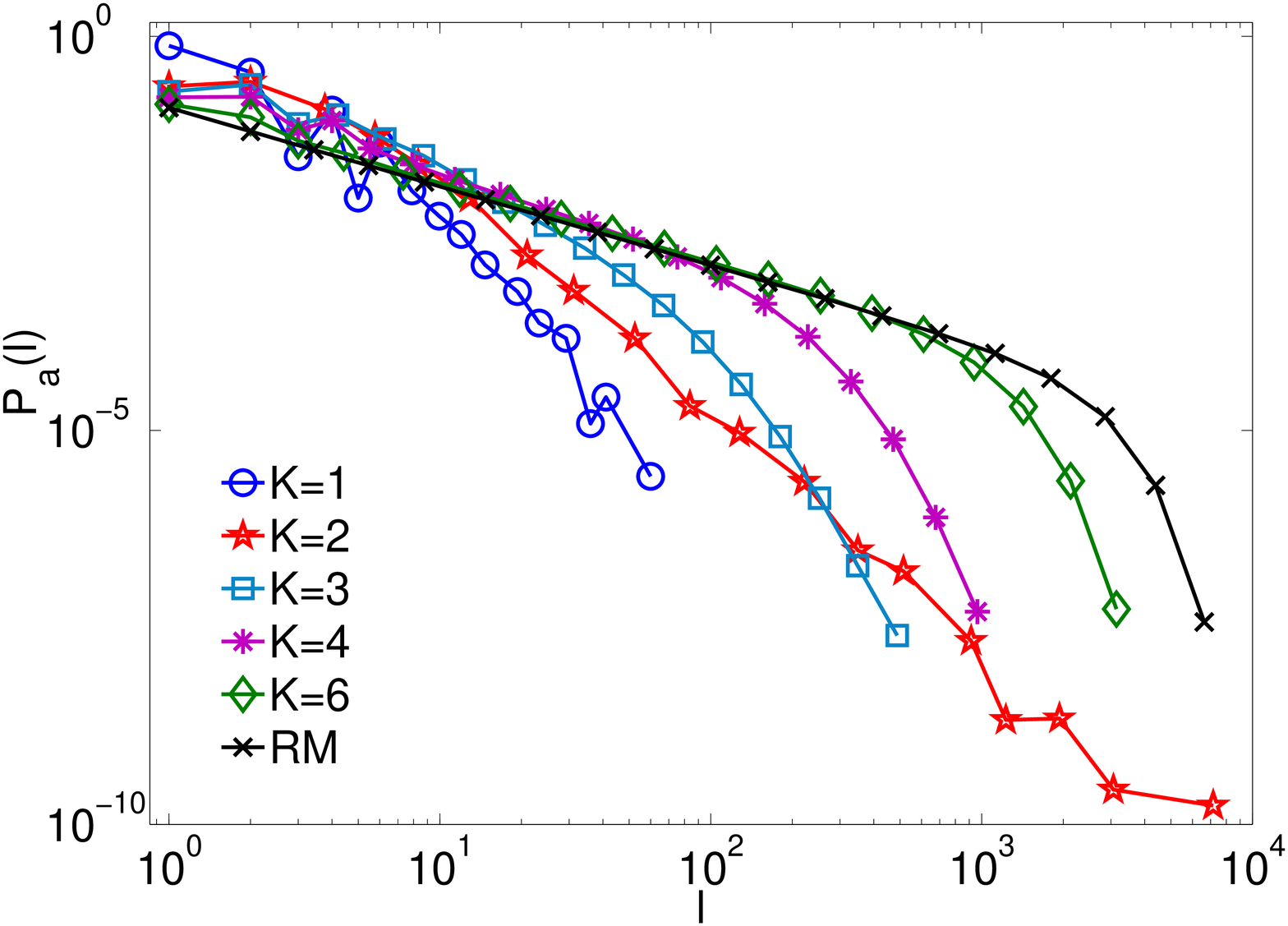,width=1\linewidth,clip=} \\
\begin{center}
(a)
\end{center}
 \epsfig{file=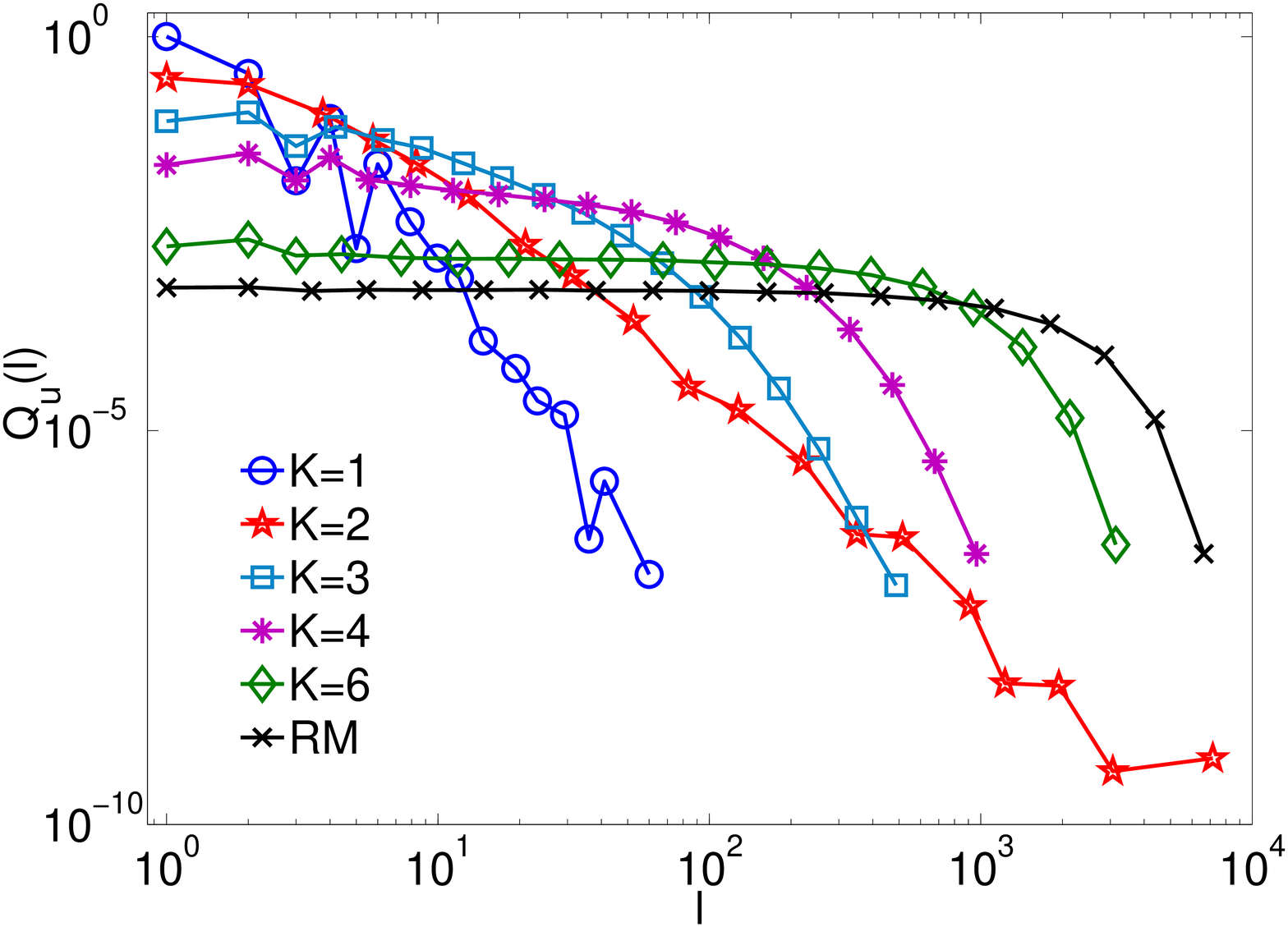,width=1\linewidth,clip=}
\begin{center}
(b)
\end{center}
\caption{(Color online.) (a) $P_{a}(l)$ and (b) $Q_{u}(l)$ for various
  $K$ and the random map (RM). $P_{a}(l)$ is a power law for all
  $K>1$. In the thermodynamic limit all $K>1$ distributions are expected 
  to have
  the same exponent, $1$. $Q_{u}(l)$ is a power law only for the
  critical $K=2$ distribution. Note that the curve for the critical
  $K=2$ ensemble does not exhibit an apparent cut-off in either case.}
 \label{LaKall}
 \end{figure}

From Eqs.~(\ref{RSALSA}) and (\ref{EEAL1}), it is straightforward to
show that $\langle b(l) \rangle_a \propto l$ for the random map.
Fig.~\ref{meanB} confirms this. Moreover, Fig.~\ref{meanB} shows that
the linear relationship also holds for large $K$ suggesting that the
qualitative differences between $P_a(l)$ and $Q_u(l)$ persist. Indeed,
using the annealed approximation, Eqs.~(\ref{RSALSA}) and
(\ref{EEAL1}) can be extended to \emph{chaotic} RBNs ($K>2$), as was
done previously for Eq.~(\ref{RSALSA}) in~\cite{bastollaparisi}. This
leads to a slightly modified finite size scaling form. For $K \geq 6$,
the expressions given in Ref.~\cite{bastollaparisi} allow us to
collapse the different curves as shown in Fig.~\ref{LaK6}. For smaller
$K$ the accessible values of $N$ are too small to apply the annealed
approximation, which is exact in the limit $N \to \infty$. This is
also confirmed by Fig.~\ref{meanB}, which indicates that the
dependence of $\langle b(l) \rangle_a$ on $l$ weakens with decreasing
$K$ for fixed $N$. This explains why for $N=22$ and $K=3$, $P_a(l)$
and $Q_u(l)$ do not show the same qualitative differences as seen for
higher values of $K$ (see Fig.~\ref{LaKall}).

\begin{figure}
\centering
\begin{tabular}{cc}
\epsfig{file=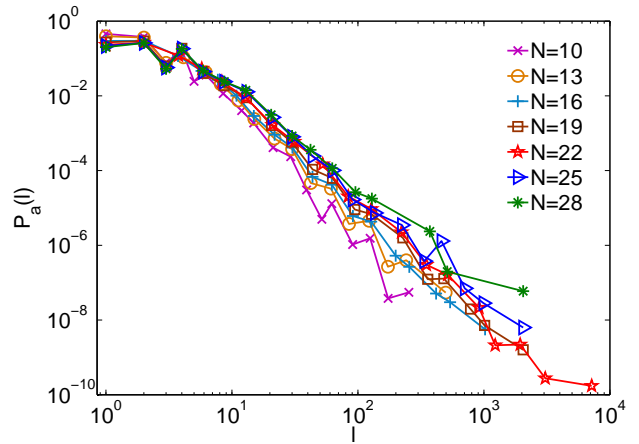,width=1\linewidth,clip=}
\end{tabular}
\caption{(Color online.) $P_{a}(l)$ for critical $K=2$ RBNs with
  various $N$. The slope of the curves varies systematically with $N$.
  Based on the results of \cite{bhattaPRL} we expect this distribution
  to be a power law with exponent $-1$ in the thermodynamic limit.
  These results are based on $5\times10^5$ realizations for $N=10-22$,
  $5\times10^4$ for $N=25$ and $5\times 10^3$ for $N=28$.}
\label{LaK2}
\end{figure}

Fig.~\ref{Wschemes} also indicates that $P_a(l)$ and $Q_u(l)$ are
statistically identical for the critical case $K=2$. As discussed in Section II (see
Eq.~(\ref{lbsamuni})), this can be directly related to the observation in
Fig.~\ref{meanB} that $\langle b(l) \rangle_a$ is basically independent of $l$
for $K=2$.  Numerical results presented in Ref.~\cite{bhattaPRL} suggest that
$Q_u(l)$ for $K=2$ decays as a power-law with an $N$-dependent exponent that
approaches $1$ in the thermodynamic limit. Our results in Fig.~\ref{LaK2}
show that the exponent of $P_a(l)$ also varies with system size. Provided that
$\langle b(l) \rangle_a$ remains independent of $l$ for $N \to \infty$, this
implies that the exponent for $P_a(l)$ also approaches $1$ in this limit.

 \begin{figure}
 \centering
 \epsfig{file=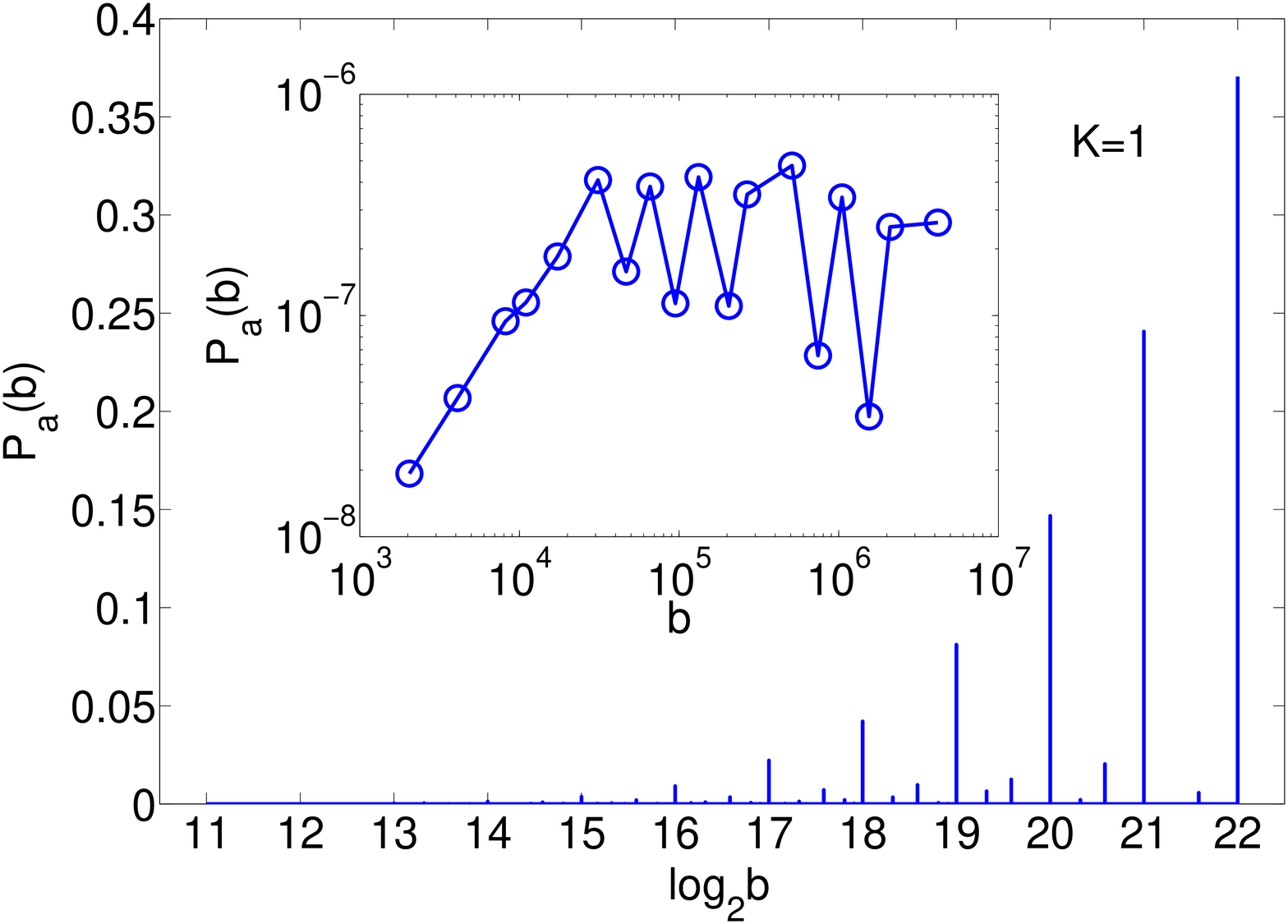,width=1\linewidth,clip=} \\
 \epsfig{file=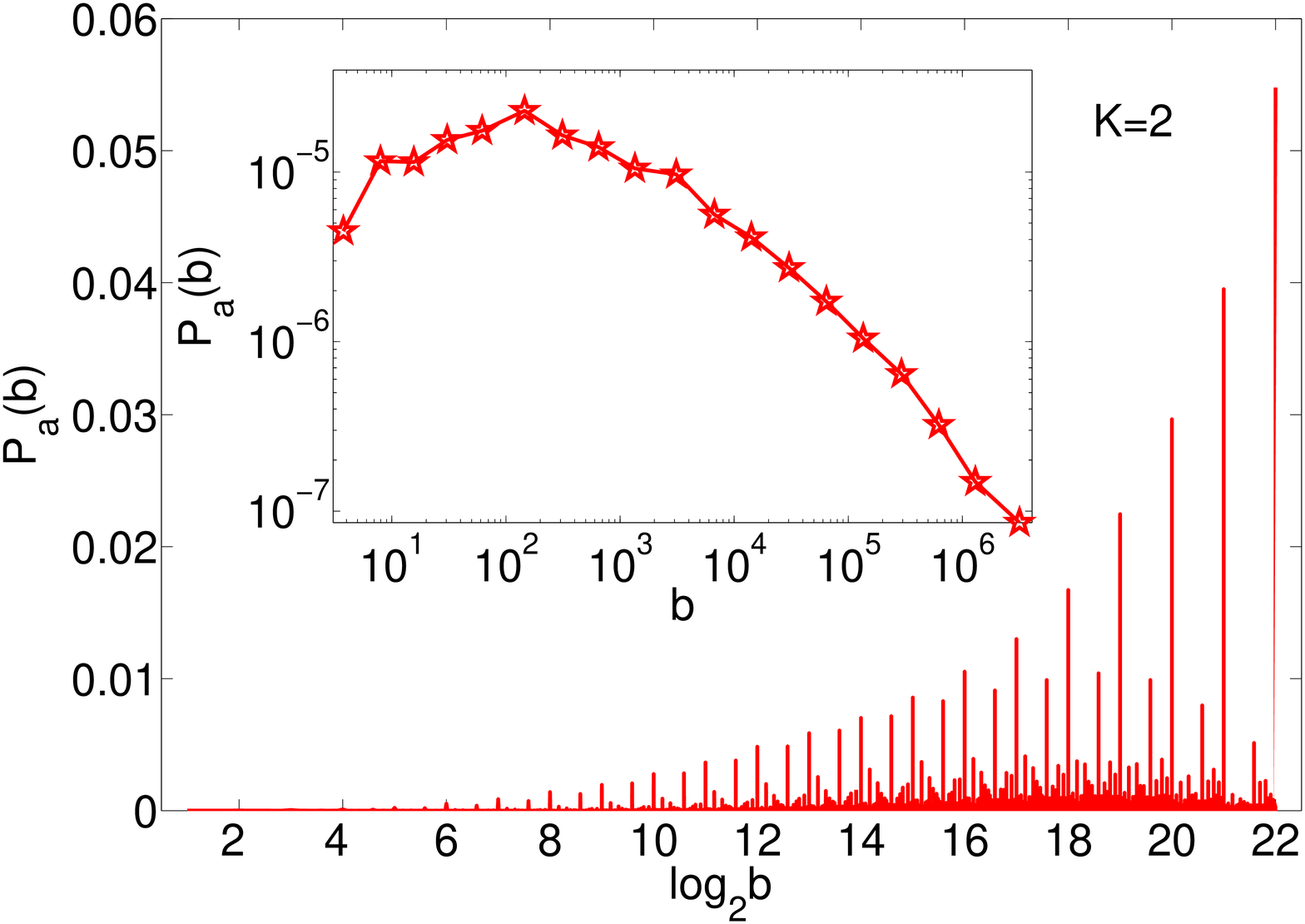,width=1\linewidth,clip=}
 \caption{(Color online.) The unbinned (main plots) and binned
   (insets) distribution of basin sizes, $P_a(b)$, for $K=1$ \& $2$.
   For $K=1$ the distribution is discrete with peaks only at integer
   multiples of powers of 2 as found in \cite{flyvbjerg1988esk}. The
   distribution for $K=2$ shares \emph{some} of this special structure
   -- it is dominated by, but not entirely restricted to, integer
   multiples of powers of 2.}
 \label{unBinnedBasins}
 \end{figure}

Consequently, the difference between the state spaces of chaotic and critical
RBNs lies \emph{not} in the distribution of cycle lengths obtained by exact
enumeration, $P_a(l)$ (Fig.~\ref{LaKall}, upper panel), which decays as a
power law for all $K>1$, but in the correlations between the attractor cycle
length and its basin size. While $\langle b(l) \rangle_a \approx$ constant for
ensembles of $K=2$ critical RBNs, $\langle b(l) \rangle_a \propto l$ for
ensembles of chaotic RBNs. The presence or absence of such correlations is
naturally captured by whether or not $Q_u(l)$ (Fig.~\ref{LaKall}, lower panel)
differs from $P_a(l)$.

Most approaches to estimate attractor length distributions in RBNs in
the past have exclusively focused on $Q_u(l)$, which is the simplest
distribution that can be obtained by randomly sampling initial states.
Indeed, the specific random sampling methods applied in
Refs.~\cite{Kauffman1969,bhattaPRL,paczuskiRBN} are simply not able to
measure $\langle b(l) \rangle_a$ or $P_a(l)$. This is the clear
advantage of the exact enumeration method applied here.

 Exact enumeration also allows us to find the sizes of the basins of
 attraction. The distribution of basin sizes is known analytically for
 $K=1$~\cite{flyvbjerg1988esk}, $K=N$~\cite{DerridaFlyvbjerg} and all
 $K$ in the chaotic phase~\cite{bastollaparisi}.
 Ref.~\cite{ShmulevichEntropyRBNPRL} used exact enumeration to
 numerically find that the basin entropy scales only for $K=2$
 critical RBNs. However, they did not present the distribution of basin sizes
 for $K=2$.

 Fig.~\ref{unBinnedBasins} shows the distribution of
 basin sizes $P_a(b) \equiv \sum_l P_a(l,b)$ for $K=1$ and $K=2$. For
 $K=1$ the basin size is given by
\begin{eqnarray}
  b &=& l2^{N-m}
 \label{K1basin}
\end{eqnarray}
where $m$ is the number of relevant elements in the
RBN~\cite{flyvbjerg1988esk}. This causes the distribution of basin
sizes for $K=1$ to be discrete and peaked at integer multiples of
powers of two. In the lower panel of Fig.~\ref{unBinnedBasins} we see
that the distribution for $K=2$ shares \emph{some} of this special
structure -- it is dominated by, but not entirely restricted to,
integer multiples of powers of 2. This is most likely because the
structure of $K=2$ relevant components is similar to those of $K=1$ as
discussed in~\cite{DrosselRBNreview2008}. Approximating $P_{a}(b)$ for $K=2$ by a 
continuous distribution is shown in the inset of Fig.~\ref{unBinnedBasins}.


\section{Conclusion}

The full enumeration of the state space for each 
realization of an RBN presented here allows us to estimate the distribution 
of attractor cycle lengths or basin sizes mimicking different sampling 
procedures and weighting schemes. The unbiased distribution of attractor
lengths, obtained by weighting all attractors in the ensemble equally, is 
very different from the cycle
length distribution of attractors reached from a randomly chosen initial
state. Yet, for the critical case $K=2$ both distributions are statistical
identical. This directly implies that the random sampling procedure applied
in Ref.~\cite{bhattaPRL} for $K=2$ and large $N$ indeed allows one to obtain
a reliable estimate for the unbiased distribution of attractor cycle lengths.
The comparison of the different weighting schemes also shows that 
the fraction of attractors of a given length in a given RBN is statistically 
independent of the total number of attractors in that RBN for all $K>1$.

In addition, our findings show that the existence of a power-law decay
in the unbiased distribution of attractor lengths is \emph{not} an
indicator of the criticality of an RBN (as defined in the
introduction), since this distribution also decays as a power law for
all RBNs in the chaotic phase ($K>2$). Thus, the difference between
the state spaces of chaotic and critical RBNs lies not in the unbiased
distribution of attractor cycle lengths, but in the correlations
between the attractor cycle length and the sizes the basins that these
attractors drain. The typically applied random sampling procedure
naturally captures this correlation between the attractor length and
the basin size. This correlation is, however, only made explicit by
measuring the joint probability of attractor length and basin size,
which is accomplished by a full enumeration of the state space of an
RBN. Such an enumeration scheme allows one in particular to obtain the
distribution of basin sizes.
For K=2, we find that the distribution is strongly peaked at integer
multiples of powers of 2. In a more general setting, this enumeration
scheme could also prove useful to investigate correlations between
attractor cycle length and basin size in systems like cellular
automata~\cite{Wolfram1983, Shreim2007CA}, discrete dynamical
mappings~\cite{kyriakopoulos2007dnr} and multi-stable dynamical
systems~\cite{feudel96,feudel97,feudel03}, or in models of genetic
regulatory
networks~\cite{albert2003dros,samal2008rnc,davidich2008bnm}.

\section{Acknowledgments}
We thank P. Grassberger for helpful discussions and useful comments. This work
was partially supported by NSERC.

\bibliography{References}

\end{document}